\documentclass[twocolumn,showpacs,preprintnumbers,amsmath,amssymb]{revtex4}

\usepackage{graphicx}
\usepackage{dcolumn}
\usepackage{bm}

\newcommand {\be}{\begin{equation}}
\newcommand {\ee}{\end{equation}}
\newcommand {\bea}{\begin{eqnarray}}
\newcommand {\eea}{\end{eqnarray}}
\newcommand {\bem}{\begin{displaymath}}
\newcommand {\eem}{\end{displaymath}}
\begin{document}

\preprint{ }

\title{ Magnetization Losses in Multifilament Coated Superconductors.}
\author{ G. A. Levin, P. N. Barnes}
\affiliation{Air Force Research Laboratory, Propulsion Directorate, 1950 Fifth Street, Bldg. 450, Wright-Patterson Air Force Base, OH 45433}
\author{ N. Amemiya, S. Kasai, K. Yoda, Z. Jiang}
\affiliation{Faculty of Engineering, Yokohama National University,
79-5 Tokiwadai, Hodogaya, Yokohama 240-8501, Japan}

\date{\today}

\begin{abstract}
We report the results of a study of the magnetization losses in experimental multifilament, as well as control (uniform), coated superconductors exposed to time-varying magnetic field of various frequencies. Both the hysteresis loss, proportional to the sweep rate of the applied magnetic field, and the coupling loss, proportional to the square of the sweep rate, have been observed. A scaling is found that allows us to quantify each of these contributions and extrapolate the results of the experiment beyond the envelope of accessible field amplitude and frequency. The combined loss in the multifilament conductor is reduced by about $90\% $ in comparison with the uniform conductor at full field penetration at sweep rate as high as 3T/s. 
\end{abstract}
\maketitle

%
\begin{figure*}
\includegraphics{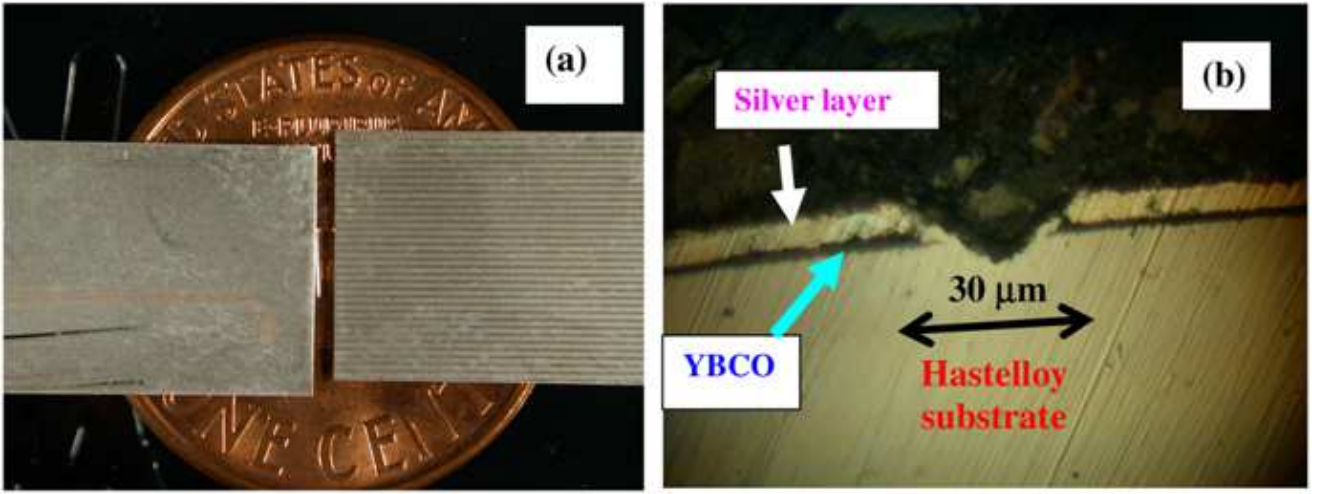} 
\caption{\label{fig:epsart} {\bf (a)} One cm wide control (uniform) coated superconductor and 33-filament striated sample shown side by side. The top visible layer is silver. {\bf (b)} Microphotograph of the cross-section of the striated sample in the groove area. The Hastelloy substrate, YBCO and silver layers are indicated.}
\end{figure*}
\section{\label{sec:level1}Introduction\protect}
Coated superconductors are currently produced by depositing a $~1-2 \mu$m thin film of $Y_1Ba_2Cu_3O_{6+x}$ ($YBCO$) on $~100\;\mu m$ thick buffered metal substrate in the form of a thin tape. The superconducting film is subsequently covered by protective silver layer\cite{Larbalestier}. The width of the developed coated conductors is 1 cm with plans to reduce it to 4 mm. When such a wide superconducting tape is exposed to a transverse time-varying magnetic field, a large amount of heat is generated\cite{Carr,Tinkham}.

A significant reduction of magnetization losses in coated conductors is a prerequisite for their use in ac power applications, such as transformers, generators and motors. It is also important for such a modification to be compatible with current techniques of manufacturing the coated conductors. Carr and Oberly\cite{CO} have proposed that magnetization losses can be reduced if the superconducting layer is divided into many parallel superconducting stripes segregated by non-superconducting resistive barriers and the tape is twisted.  Cobb et al.\cite{Cobb} have demonstrated the validity of this approach on small samples of YBCO films deposited on $LaAlO_3$  single crystal substrate. 
However, because of the small size of the samples and insulating substrate, coupling losses due to currents induced between the superconducting stripes\cite{Carr} could not be detected. 

Here, we report the results of the magnetization losses measurements in actual multifilamentary YBCO coated conductors. Two $1\times 10\; cm$ samples of YBCO coated conductors were subjected to a magnetic field normal to the wide face of the tape and varying at different frequencies $f$ in the range $11-170\; Hz$ with the magnetic induction amplitude $B$ up to $70\; mT$. The longer samples and measurements at several frequencies enabled us to study both hysteresis and coupling losses\cite{Amemiya}. We found a scaling that allows to separate and quantify the contributions of the two types of losses. The measured magnetization losses are compared with theoretical estimates and some of the implications of our findings for the prospects of coated superconductors for power applications are discussed. 
\section{\label{sec:level1}Experiment and Analysis\protect}

The samples were cut from a longer tape of YBCO coated conductor with Hastelloy substrate provided by SuperPower, Inc.\cite{S} The multifilamentary sample used for the loss measurements was made by dividing the uniform tape into 20 stripes by laser ablation\cite{Cobb,Hix}. In  Fig. 1(a) a similar sample, only with greater number of stripes (33),  is shown next to the control sample. In Fig 1(b) the cross-section of a typical groove which has a characteristic triangular shape is shown. The grooves less than $50\mu m$ wide separating the superconducting stripes were cut through the silver protective layer, $YBCO$ and buffer layers, and extended $~5-10\mu m$ into the substrate. In the groove area the aforementioned material was evaporated. The electrical connection between the YBCO layer and the substrate is predominantly provided by Hastelloy that was melted and splashed across the walls of the grove during ablation. More details about the properties of the grooves and their dependence on the processing parameters can be found in Ref.\cite{Hix}. 
\begin{figure*}
\includegraphics{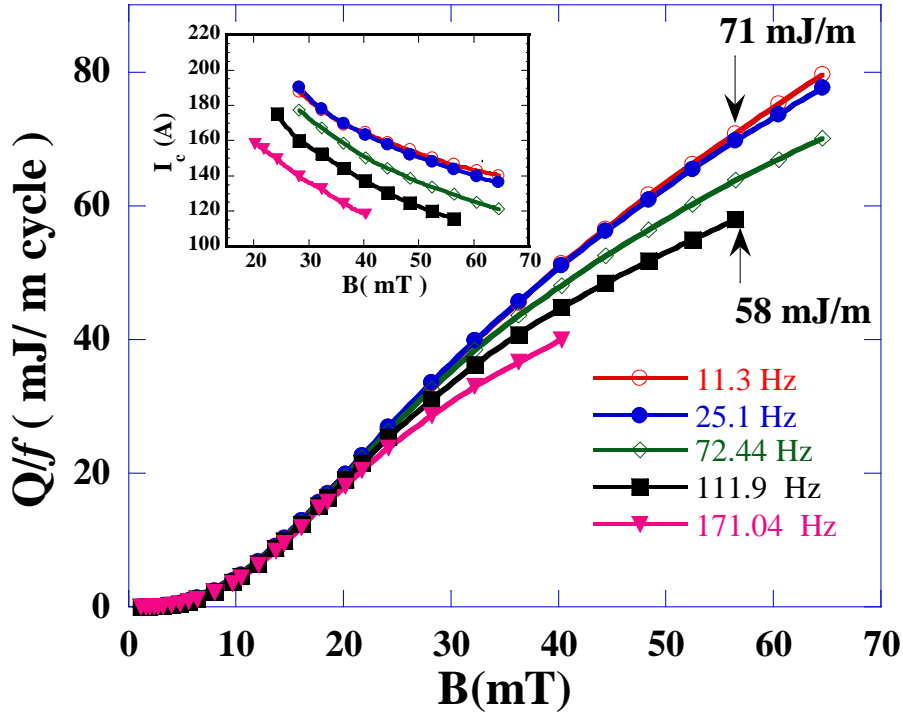} 
\caption{\label{fig:epsart} Energy loss in the uniform sample per unit length, per cycle, $Q/f$,  at five different frequencies, from $11.3$ to $171\; Hz$, plotted vs  amplitude of the applied magnetic field. The arrows indicate the values of losses at $B=57\; mT$ and two frequencies, $11.3$ and $111.9\; Hz$.  The inset shows the bulk average critical current obtained from the loss data by inverting Eq. (1).}
\end{figure*}

The losses were measured using a linked pick-up coil as described in Refs.\cite{Jiang,Amemiya}. The entire system was cooled in liquid nitrogen. A sample was placed inside the bore of an ac dipole magnet that generated a magnetic field whose frequency was varied from $11.3$ to $171.0\; Hz$.
\begin{figure*}
\includegraphics{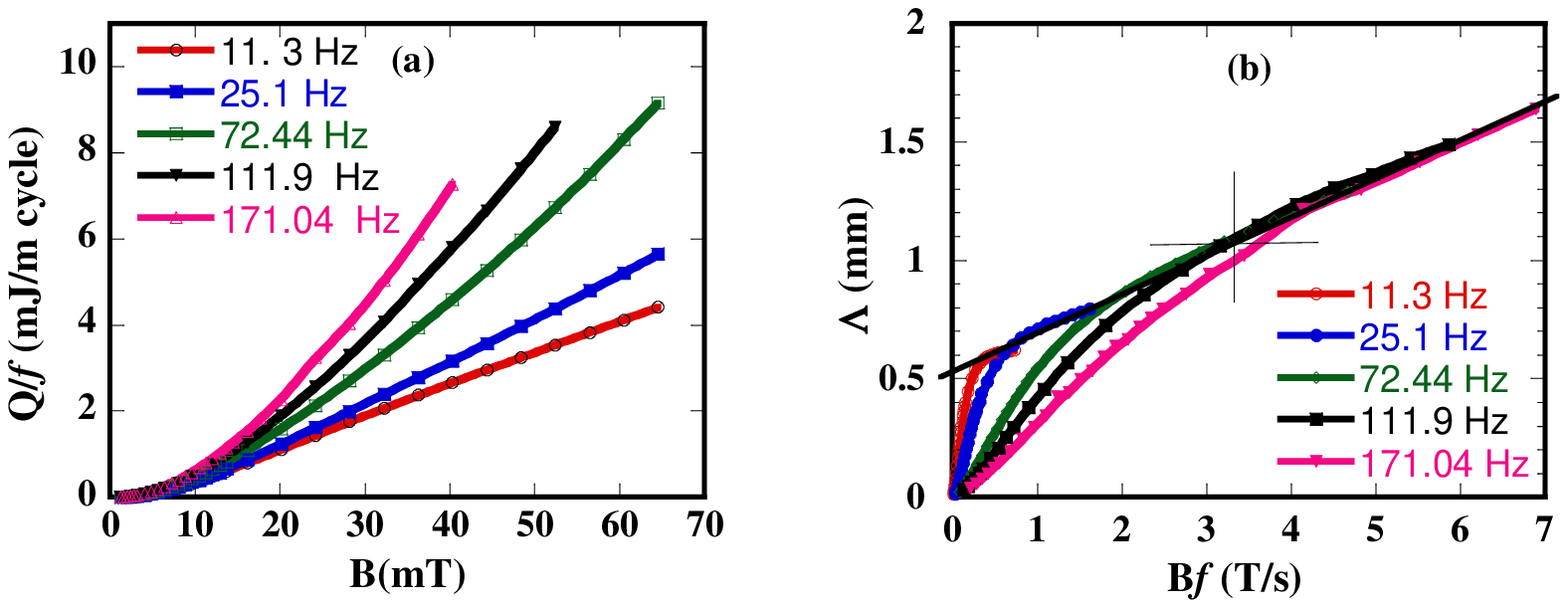} 
\caption{\label{fig:epsart} {\bf (a)}The energy loss per cycle, $Q/f$, in the 20-filament sample, plotted vs amplitude of the applied magnetic field. {\bf (b)}  The specific loss  $\Lambda\equiv Q/I_cBf$ in $mm$ vs sweep rate. The plot includes all the data from Fig. 3(a). The loss data were normalized by a constant value of measured transport current  $I_c=110\;A$. The straight line is an approximate linear fit corresponding to Eq. (5). Crossed lines in the center indicate the break-even point, where the coupling losses are equal to hysteresis loss. }
\end{figure*}

In Figure 2 the losses in the uniform sample at different frequencies are shown. 
For comparison, we use the analytical result for the hysteresis loss\cite{Brandt}: 
\be
Q_h\approx J_cW^2(B-B_c)f;\;\;B>>B_c.
\ee
Here $Q_h$ is the hysteresis power loss per unit length, $J_c$ is the critical current per unit width, and $W$ is the width of the sample. $B_c=\mu_0 J_c\ln 4/\pi$ and $\mu_0=4\pi\times 10^{-7}\; H/m$ is the magnetic permeability of vacuum. 

The energy loss per cycle {\it decreases} with increasing frequency. The system of vortices undergoes a cyclic change in alternating magnetic field. Over a finite time the vortices cannot completely relax to the lowest energy configuration and the average over the cycle pinning energy is lower than in the steady magnetic field. This leads to the reduction of the critical current and the corresponding reduction of the per cycle hysteresis loss with frequency. 
Inverting Eq.(1) within its range of applicability we obtained the critical current $I_c(B,f)=J_cW$ shown in the inset.  The value of $I_c$ obtained from ac losses is the bulk average and is always greater than the transport critical current which is limited by the weakest section of the conductor. Unfortunately, the transport current could not be reproducibly measured in the uniform sample because it was accidentally destroyed during the measurement. The manufacturer provided the self-field value of $I_c =132\;A$ for that particular section of the tape. 

In a multifilamentary sample the hysteresis loss is proportional to the width of an individual stripe\cite{CO,Cobb,Mawatari}:
\be
Q_h^{st}=I_cW_n(B-B_c^{\prime})f.
\ee
Here $W_n$ is the width of the stripe, and $I_c$ is the total critical current. In our sample the distance between the centers of the neighboring grooves is $0.5\; mm$, while the width of the superconducting stripes $W_n\approx 0.45\; mm$. 

Figure 3(a) shows the loss per cycle at five frequencies in the multifilament sample. The losses {\it increase} with frequency, while the field dependence, which is essentially linear at the lowest frequency, acquires a clear curvature at higher frequencies. The loss that has quadratic dependence on the sweep rate $Bf$ is the coupling loss\cite{Carr}. The eddy current losses in the silver layer and substrate are negligible because the silver cap layer is striated along with the $YBCO$ layer. The average electric field directed perpendicular to the stripes, $E_{\perp}\sim BfL$. Here $L$ is the length of the sample or, in a long twisted conductor, half of the twist pitch. The power per unit length dissipated as the result of the current passing through the metal connecting the stripes is\cite{CO} 
\be
Q_n^{st}\sim \rho^{-1}|E_{\perp}|^2d_nW=2\frac{(BfL)^2}{\rho}d_nW.
\ee
Here $d_n$ is the thickness of the normal metal layer 
and $\rho$ is the resistivity. 
The combined magnetization loss can be defined as follows:
\be
Q=Q_h^{st}+Q_n^{st}\equiv\Lambda I_c Bf,
\ee                                       
where $\Lambda$ is defined as the power loss per unit length, per unit of the critical current, per unit of the sweep rate. It has the dimensionality of length.
Comparison with Eqs.(2) and (3) suggests that in the limit $B>>B_c^{\prime}$, 
\be
\Lambda = \lambda_1+\lambda_2 Bf,
\ee                                 
where 
\be
\lambda_1\approx W_n;\;\; \lambda_2 = 2\frac{L^2}{\rho I_c}d_nW.
\ee        
Thus, these two parameters are independent of the sweep rate and characterize the conductor itself. One of them determines the hysteresis loss and the other the coupling loss\cite{Tinkham}. Moreover, $\Lambda$ is a more objective criterion of how successful the striation is in reducing the net loss than the value of the loss itself, since it accounts for the reduction of the critical current by striation. 
\begin{figure*}
\includegraphics{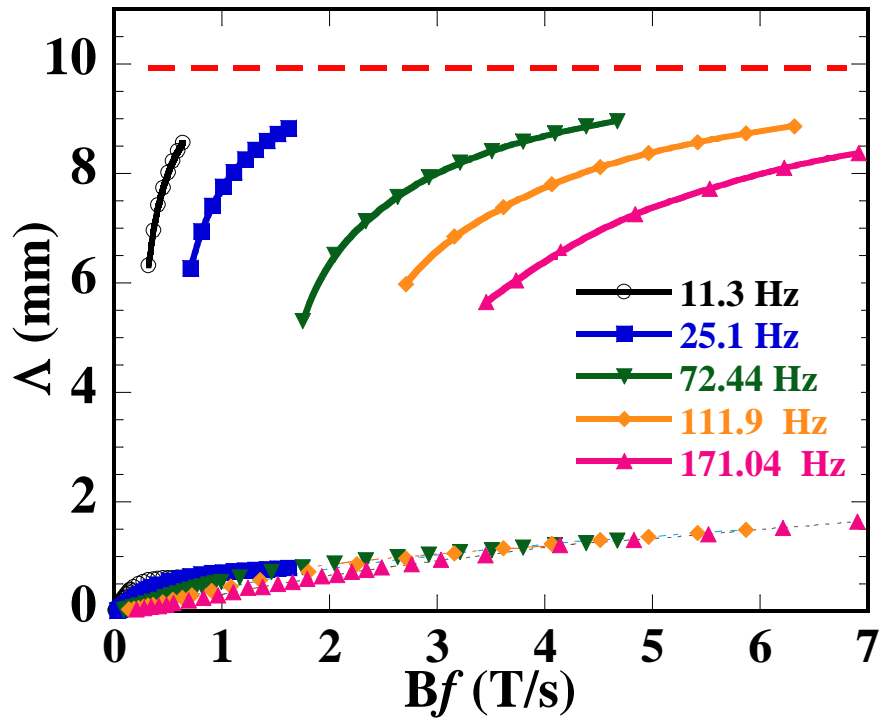} 
\caption{\label{fig:epsart} The data for uniform and multifilament samples are shown together. The plot includes the data from Fig.2 presented as $\Lambda =Q/I_cBf$ and from Fig. 3(b). For the uniform sample we have used the values of the critical current shown in the inset to Fig.2. }
\end{figure*}

The scaling defined by Eq. (5) is demonstrated in Fig. 3(b). It shows the same data as in Fig.3(a), but presented as $\Lambda$ vs sweep rate. The transport critical current $I_c=110\;A$ was measured directly in this sample\cite{Amemiya}.
The linear fit shown by the solid line allows us to determine $ \lambda_1 =0.52\;mm $ and $\lambda_2 =0.155\;mm\;s/T$. Equation (5) can also be rewritten as
\be
\Lambda=\lambda_1\left (1+\frac{Bf}{\cal R}\right );\;\; 
{\cal R} =  \frac{\lambda_1}{\lambda_2}=\frac{1}{2}\frac{W_n}{W}\frac{\rho I_c }{d_n L^2}.
\ee
Here $\cal R$ is the {\it break-even sweep rate} at which the coupling loss is equal to the hysteresis loss. For our sample, ${\cal R} \approx 3.3\; T/s$. 

Resistivity in the direction perpendicular to the grooves is that of the Hastelloy substrate, $\rho\approx 130\mu\Omega$ cm. Using Eq.(7) we can estimate the break-even rate $\cal R$. Taking $W_n\approx 0.5\; mm$,  $W = 10$ mm, $L = 10$ cm,  the experimental values\cite{Amemiya}  of the sheet resistance $\rho /d_n \approx 13 m\Omega$ and $I_c = 110\; A$, we obtain 
\be
{\cal R}\approx  \frac{W_n}{2W}\frac{\rho  I_c }{d_n L^2}\approx 3.6\; T/s.
\ee
This is not a bad agreement with the experimental value (3.3 T/s), considering that Eq. (7) is practically an order-of-magnitude estimate. 

In Fig. 4 the losses for striated and uniform samples are presented together. The data for the uniform sample are normalized to the field and frequency dependent critical current, shown in the inset to Fig.2. Given the small coupling losses in the uniform sample, $\Lambda\approx W(1-B_c/B)$ with $B_c\sim 10 \;mT$. Note that this is a conservative comparison because the losses in the uniform sample were normalized by the bulk average critical current, which is always greater than the transport critical current we have used to normalize losses in the multifilament sample. 

An objective criterion of the effectiveness of striation in reducing losses at full field penetration is the ratio $\Lambda /W$. As Figs. 3(b) and 4 demonstrate, it extrapolates to approximately $5\%$ in the limit of low sweep rate and increases to $10\%$ at the break-even rate. Since the coefficients $\cal R$ and $\lambda_1$ are known, we can extrapolate the value of loss outside the tested range of parameters. The striation will lose its effectiveness, in comparison with the uniform sample, at the sweep rate about $60\; T/s$ when the extrapolated $\Lambda (Bf) = 10\;mm$. 
\section{\label{sec:level1}Discussion and Summary\protect}

Let us consider the implications of these results for the prospects of coated superconductors in ac power applications. The specifications for future conductors will likely be defined by three criteria: the level of acceptable power loss per unit length, $Q_{goal}$, at an operating sweep rate $(B_{\perp}f)_0$, and the critical current $I_c$ ($B_{\perp}$  is the normal to the face of the tape component of the magnetic induction). 
Then, according to Eq.(4),  these parameters  determine the value of 
\be
\Lambda=\frac{Q_{goal}}{I_c(B_{\perp}f)_{0}}.
\ee
In the optimally designed conductor, under most circumstances, the hysteresis loss should be approximately equal to the coupling loss at the operating rate, which means   
\be
{\cal R}\approx (B_{\perp}f)_0.
\ee
Conditions (9) and (10) determine the required intrinsic parameters $\lambda_1$ and $\lambda_2$: 
\be
\lambda_1\approx\frac{\Lambda}{2};\;\; \lambda_2\approx \frac{\lambda_1}{(B_{\perp}f)_0}.
\ee

As an example, consider a fairly ambitious goal of achieving the level of loss per meter length per amp of critical current $Q_{goal}/I_c=1\;mW/Am$ at the operating sweep rate $(B_{\perp}f)_0 =10 \;T/s$. These conditions demand $\Lambda = 0.1 mm$ and, correspondingly, the hysteresis length $\lambda_1\approx 50\mu m$. In comparison to sample used here, this is a ten-fold increase in the number density of stripes $-$ from $20$ to $200\;stripes/cm$. The coupling constant $\lambda_2$ has to be reduced to $5\times 10^{-3} mm\;s/T$, which is about $30$ times smaller than the experimental value. 

If we take Eq.(6) at face value, the main parameter that can be substantially changed is the average resistivity of the normal metal. To increase the effective resistivity by more than an order of magnitude over that of Hastelloy  will probably require a non-metallic, insulating medium between the stripes and  an unbroken insulating buffer between $YBCO$ and metal substrate. The protective layer ($Ag$ or $Cu$) has to be striated as well, as done in our sample. 

In summary, we have tested a multifilament design of YBCO coated conductor that at sweep rate $\sim 3\;T/s$ allows reduction of magnetization loss by an order of magnitude in comparison with the uniform sample fully penetrated by the magnetic field. Analysis presented here shows that even though most ac magnets available for such experiments can achieve the sweep rate of only a few $T/s$, the scaling shown in Fig. 3(b) can determine the intrinsic characteristics of a tested conductor ($\lambda_1$ and $\lambda_2$) and predict what the losses will be well outside the envelope of accessible values of the field amplitude $B$ and frequency $f$. This type of tests can guide the development of new conductor designs needed to help reduce the magnetization losses by another one or two orders of magnitude in order to allow the implementation of coated superconductors in demanding ac power applications.
 
\begin{acknowledgments}
We would like to thank C. E. Oberly and M. Sumption for many enlightening conversations and discussions and J. Murphy and N. Yust for technical help.
This work was supported by the Air Force Office of Scientific Research under contract No. AOARD-03-4031. G.A.L. was supported by the National Research Council Senior Research Associateship Award at the Air Force Research Laboratory.
\end{acknowledgments}

\newpage

\newpage

\end{document}